\documentclass{article}
\usepackage{spconf,amsmath,graphicx}
\usepackage{microtype}
\usepackage{booktabs}
\usepackage{multirow}
\usepackage{array}
\usepackage{booktabs}
\usepackage{xcolor}
\usepackage{colortbl}
\usepackage{caption}
\usepackage{float}
\usepackage[hyphens]{url}

\usepackage{balance}
\usepackage{enumitem}
\usepackage{graphicx}
\setlist{nosep, leftmargin=14pt}
\usepackage[compatibility=false]{caption}
\usepackage{booktabs}
\usepackage{xcolor}
\usepackage{multicol}
\usepackage{tabularx}

\usepackage{mwe} 

\title{Synergy vs. Noise: Performance-Guided Multimodal Fusion For Biochemical Recurrence-Free Survival in Prostate Cancer}
\name{\begin{tabular}[t]{c}
    Seth Alain Chang$^{1\star}$, Muhammad Mueez Amjad$^{1\star}$, \\
    Noorul Wahab$^{1}$, Ethar Alzaid$^{1}$, Nasir Rajpoot$^{1}$, Adam Shephard$^{1}$
\end{tabular}}

\address{$^{1}$Tissue Image Analytics Centre, Department of Computer Science, University of Warwick, UK
\\ $^{\star}$Joint first authors contributed equally}
\begin{document}
%
\maketitle
\begin{abstract}
Multimodal deep learning (MDL) has emerged as a transformative approach in computational pathology. By integrating complementary information from multiple data sources, MDL models have demonstrated superior predictive performance across diverse clinical tasks compared to unimodal models. However, the assumption that combining modalities inherently improves performance remains largely unexamined. We hypothesise that multimodal gains depend critically on the predictive quality of individual modalities, and that integrating weak modalities may introduce noise rather than complementary information. We test this hypothesis on a prostate cancer dataset with histopathology, radiology, and clinical data to predict time-to-biochemical recurrence. Our results confirm that combining high-performing modalities yield superior performance compared to unimodal approaches. However, integrating a poor-performing modality with other higher-performing modalities degrades predictive accuracy. These findings demonstrate that multimodal benefit requires selective, performance-guided integration rather than indiscriminate modality combination, with implications for MDL design across computational pathology and medical imaging.
\end{abstract}

\section{Introduction}
Multimodal deep learning (MDL) integrates heterogeneous data sources, such as whole-slide histopathology images, molecular profiles, imaging data, and clinical variables for clinical tasks. MDL models capture complementary patterns across modalities that improve prognostic accuracy across various tasks \cite{MDL_summary}. This has led to mean increases in area under the receiver operating characteristic curve (AUROC) of 6.4\% over their single-modality counterparts \cite{MDL_AUC}, with certain MDL models reaching clinical-grade performance on par with human experts \cite{MDL_human_experts}. This multimodal integration has enabled advances across diverse tasks, including disease subtyping, survival prediction, and biomarker discovery \cite{MDL_diverse_tasks}. These advances are paving the way for clinical-grade AI systems that approach human expert-level performance, positioning MDL as a critical tool for a variety of clinical tasks. 

Contemporary fusion methodologies in medical imaging include early, intermediate, and late fusion \cite{MDL_methods}. These strategies operate at distinct stages of the analysis pipeline: early fusion combines raw input data before feature extraction, intermediate fusion integrates features extracted from individual modalities, and late fusion combines predictions from separate unimodal models \cite{fusion_strategies}. A recent systematic review identified intermediate fusion as the most promising approach in computational pathology due to its ability to effectively combine modality-specific features during training \cite{Intermediate_Fusion_Benefits}. 

Intermediate fusion can be implemented through two primary approaches: marginal intermediate fusion and joint intermediate fusion \cite{MDL_human_experts}. Marginal intermediate fusion involves combining features from different modalities without cross-modal interaction during combination through mechanisms such as element-wise operations (e.g. summation, multiplication), concatenation, or more complex tensor interactions, such as bilinear fusion \cite{IF_bilinear_fusion}. Joint intermediate fusion allows features to interact across modalities during combination, potentially capturing cross-modal dependencies and revealing more nuanced relationships between modalities that may be missed by simple combination methods. Joint intermediate fusion can be implemented in various ways, including architectures with self-attention and cross-attention layers \cite{IF_CA_SA}, and mechanisms that dynamically weight modality interactions \cite{IF_weighted_modality}.

Despite the success of multimodal fusion in computational pathology, the assumption that integrating additional modalities universally improves performance warrants critical examination. Recent studies in other fields have revealed that multimodal models may, in some cases, perform worse than their unimodal counterparts, particularly when modalities possess limited discriminative information and substantial noise \cite{fusion_challenges}.  To explore this, we systematically evaluated how integrating modalities with varying predictive capabilities affects model performance, hypothesising that high-performing modalities enhance performance while poor-performing ones degrade it.

Our main contributions are summarised as follows:
\begin{enumerate}
    \item We investigate the hypothesis that integrating a low-performing modality with a high-performing one can degrade predictive accuracy for predicting time-to-biochemical recurrence (BCR) in prostate cancer.
    \item We show that fusing high-performing modalities yields a synergistic benefit in prognostic performance, while indiscriminate fusion with weak modalities can be detrimental.
    \item We advocate a performance-guided, selective approach to multimodal fusion in computational pathology, challenging the assumption that more modalities always help.
    \item We implement a joint intermediate fusion architecture with cross-attention and self-attention to capture inter- and intra-modal dependencies, achieving superior  accuracy.
\end{enumerate}

\section{Materials \& Methods}

\subsection{Study Data}
We utilised data from Task 1 of the CHIMERA Challenge \cite{CHIMERA}, which integrates histopathology, radiology, and clinical data to predict time-to-BCR in prostate cancer patients following radical prostatectomy. The dataset comprises 95 patient cases, each including: 
\begin{enumerate}
    \item \textbf{Histopathology:} H\&E-stained Whole Slide Images (WSIs) of prostatectomy specimens (up to 12 slides per patient), scanned using a 3DHISTECH PANNORAMIC 1000 (0.5 µm/pixel).
    
    \item \textbf{Radiology:} Preoperative imaging from Siemens 3T MRI scanners, including:
        \begin{itemize}    
            \item T2-Weighted (T2W) MRI
            \item Apparent Diffusion Coefficient (ADC) maps
            \item High B-Value (HBV) Diffusion-Weighted Imaging (DWI)
            \item Prostate gland segmentation masks
        \end{itemize}
        
    \item \textbf{Clinical Metadata:} Patient-specific clinical features and outcome annotations (JSON format):
        \begin{itemize}[noitemsep,topsep=0pt,partopsep=0pt]
        \item \textbf{Demographics:} Age at prostatectomy
        \item \textbf{Grading:} Primary/Secondary/Tertiary Gleason, ISUP grade
        \item \textbf{Biomarkers:} PSA (pre-surgery, at recurrence), BCR status
        \item \textbf{Pathology:} pT-staging, invasion markers (LN, CP, SV, LV), surgical margins
        \item \textbf{History:} Earlier therapy, time to last follow-up/BCR
        \end{itemize}
\end{enumerate}

This task involves modelling time-to-BCR as a survival analysis problem. The dataset exhibits right-censoring, where not all patients experienced recurrence within the follow-up period. Specifically, 27 patients (28.4\%) had confirmed BCR events, while 68 patients (71.6\%) were censored without observed recurrence.

\subsection{Approach}
Our analysis proceeded in three stages. First, we independently assessed each of the three modalities to quantify their baseline predictive performance for the clinical task. Second, we systematically evaluated all three possible pairwise combinations. Finally, we assessed the performance of a model combining all three modalities. 

\subsubsection{Model Architecture}
All models were implemented as multilayer perceptrons (MLPs) with three hidden layers, ReLU activations, and a normalisation layer. Hyperparameters (layer sizes, dropout rates) were tuned for each configuration. Models were trained using the Adam optimiser and a Cox proportional hazards loss function, and early stopping was applied to prevent overfitting. Performance was evaluated using the concordance index (C-Index) \cite{C_Index} on hold-out test sets comprising 20\% of the data within each fold of a 5-fold cross-validation scheme, repeated 10 times (with random seeds) to ensure robust estimates.

\subsubsection{Unimodal Models}
We conducted a comprehensive evaluation of pretrained foundation models to determine the most effective feature extraction methods for histopathology and radiology data. \textbf{For histopathology images}, we tested different slide-level foundation models (TITAN \cite{TITAN}, CHIEF \cite{CHIEF}, PRISM \cite{PRISM}, MADELEINE \cite{MADELEINE} implemented using the TRIDENT repository \cite{TRIDENT_1}\cite{TRIDENT_2}), with random selection employed when multiple slides were available per patient. \textbf{For radiology images}, we leveraged a general-purpose 3D medical imaging encoder, specifically MedicalNet \cite{MedicalNet}, applying it to T2-weighted sequences. T2-weighted imaging offers superior contrast for lesion detection and is less affected by motion or diffusion artifacts compared to HBV and ADC sequences. For further comparison against a more interpretable pipeline, we also extracted handcrafted radiomic features (n=120), based on shape, intensity and texture \cite{Radiomics}.\textbf{ For clinical data}, numerical and categorical variables were manually processed into feature vectors. Categorical variables were one-hot encoded, with each feature vector component representing either a numerical field or a specific category.

\begin{figure*}[t]
    \centering
    \includegraphics[width=14cm, height=8cm]{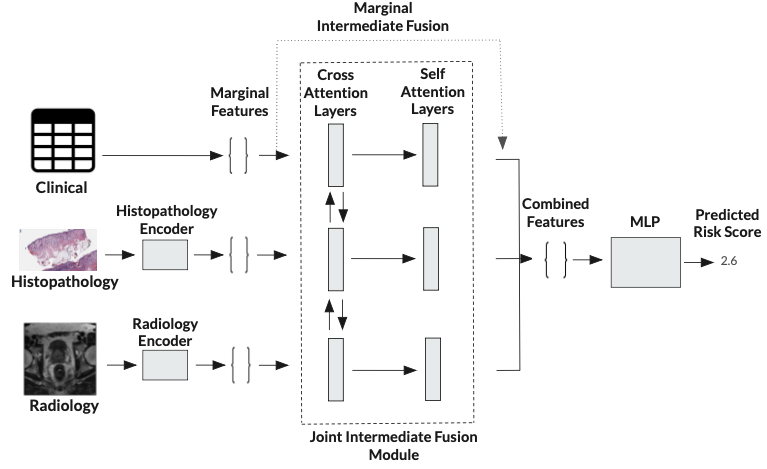}
    \caption{Overview of multimodal fusion strategies for predicting biochemical recurrence-free survival in prostate cancer. (a) Marginal intermediate fusion (dotted line): modality-specific features are concatenated without cross-modal interaction before prediction.
    (b) Joint intermediate fusion: features interact through cross-attention and self-attention layers to capture inter- and intra-modal dependencies prior to prediction.}
    \label{fig:my_wide_figure}
\end{figure*}

We adopted intermediate fusion for our multimodal architecture based on theoretical considerations. Intermediate fusion addresses key limitations of alternative approaches for the clinical task of predicting time-to-BCR: early fusion would compromise the feature extraction capabilities of modality-specific foundation models by combining raw images before leveraging their domain-specific training, while late fusion would miss critical cross-modal interactions that may be essential for accurate BCR prediction. Thus, we tested two different fusion approaches: marginal and joint intermediate fusion. In both approaches, features from each modality were first standardised to the same dimension via linear projection to ensure equal representation and prevent any single modality from dominating the fusion.

\subsubsection{Marginal Intermediate Fusion}
For marginal intermediate fusion, features from each modality were encoded independently, concatenated, and passed to an MLP for prediction (see Fig. \ref{fig:my_wide_figure}).

\subsubsection{Joint Intermediate Fusion}
In our joint intermediate fusion pipeline, encoded features from each modality were processed through successive cross-attention and self-attention layers before prediction (see Fig. \ref{fig:my_wide_figure}). Cross-attention facilitates interaction between modalities, enabling features to query each other and establish correspondences before fusion. For example, histopathology features related to cellular proliferation may attend to radiology features capturing enhancement patterns indicative of similar biological processes.

After applying cross-attention across all modalities, self-attention was applied within each modality to model internal dependencies among feature components. This step helps create contextual representations where related elements can influence each other. For instance, histopathology features representing cell density may attend to those representing cell size, reflecting their joint role in cancer progression. Finally, the updated features were concatenated and passed through a predictive MLP.

\section{Results and Discussion}

\subsection{Unimodal Model Performance}

\renewcommand{\arraystretch}{1}
\begin{table}[!ht] 
\centering
\small
\begin{tabular}{>{\centering}p{2cm} >{\centering}p{2.5cm} >{\centering\arraybackslash}p{2.5cm}}
\toprule
\textbf{Patch Size and Resolution} & \textbf{Model} & \textbf{C-Index} \\
\midrule
\multirow{4}{2.5cm}{\centering\textbf{10$\times$ 256}} 
& PRISM \cite{PRISM} & 0.7759 ± 0.1088 \\
& CHIEF \cite{CHIEF} & 0.7720 ± 0.0944 \\
& TITAN \cite{TITAN} & 0.6916 ± 0.1049 \\
& MADELEINE \cite{MADELEINE} & 0.7307 ± 0.1038 \\
\midrule
\multirow{4}{2.5cm}{\centering\textbf{10$\times$ 512}} 
& PRISM \cite{PRISM} & 0.7392 ± 0.1095 \\
& CHIEF \cite{CHIEF} & 0.7380 ± 0.1121 \\
& TITAN \cite{TITAN} & 0.6906 ± 0.1125 \\
& MADELEINE \cite{MADELEINE} & 0.6498 ± 0.1275 \\
\midrule
\multirow{4}{2.5cm}{\centering\textbf{20$\times$ 256}} 
& \textbf{PRISM} \cite{PRISM}& \textbf{0.7939 ± 0.1106} \\
& CHIEF \cite{CHIEF} & 0.6717 ± 0.1240 \\
& TITAN \cite{TITAN} & 0.7418 ± 0.1101 \\
& MADELEINE \cite{MADELEINE} & 0.7065 ± 0.1035 \\
\midrule
\multirow{4}{2.5cm}{\centering\textbf{20$\times$ 512}} 
& PRISM \cite{PRISM} & 0.7594 ± 0.1100 \\
& CHIEF \cite{CHIEF} & 0.6741 ± 0.1140 \\
& TITAN \cite{TITAN} & 0.7172 ± 0.1056 \\
& MADELEINE \cite{MADELEINE} & 0.7159 ± 0.1007 \\
\bottomrule
\end{tabular}
\caption{Performance of histopathology foundation models across patch sizes and magnifications, reported as mean C-index $\pm$ standard deviation over repeated (10 repeats) 5-fold cross-validation experiments}
\label{tab:slide_configs}
\end{table}

\renewcommand{\arraystretch}{1.3}
\begin{table}[t] 
\small
\centering
\begin{tabular}{>{\centering}p{2.5cm} >{\centering\arraybackslash}p{2.5cm}}
\toprule
\textbf{Model} & \textbf{C-Index} \\
\midrule
\textbf{MedicalNet \cite{MedicalNet}} & \textbf{0.5584 ± 0.1250} \\
Radiomics \cite{Radiomics} & 0.5508 ± 0.1318 \\
\bottomrule
\end{tabular}
\caption{Performance of radiology models over repeated (10 repeats) 5-fold cross-validation experiments. We present mean C-index $\pm$ standard deviation.}
\label{tab:unimodal_models}
\end{table}

\begin{table}[!ht]
\centering
\begin{tabular}{lcc}
\toprule
\textbf{Modality} & \textbf{C-Index} \\
\midrule
\textbf{Clinical} & \textbf{0.8037 ± 0.1034} \\
Histopathology & 0.7939 ± 0.1106 \\
Radiology & 0.5584 ± 0.1250 \\
\bottomrule
\end{tabular}
\caption{Best results for each modality over repeated (10 repeats) 5-fold cross-validation experiments. We present mean C-index $\pm$ standard deviation.}
\label{tab:rankings}
\end{table}

To identify the most effective feature representations for predicting time-to-BCR, we evaluated a range of histopathology and radiology foundation models. For histopathology, we tested multiple patch sizes (256$\times$256 and 512$\times$512 pixels) across magnifications (10$\times$, 20$\times$). As shown in Table \ref{tab:slide_configs}, PRISM (20$\times$ magnification, 512$\times$512 pixels) achieved the highest perforamce (C-Index = 0.7939). For radiology, MedicalNet yielded the best results (C-Index = 0.5584), though performance remained close to random (Table \ref{tab:unimodal_models}).

Table \ref{tab:rankings} summarises the best performing models per modality. Clinical data demonstrated the strongest predictive performance (C-Index = 0.8037), followed closely by histopathology (C-Index = 0.7939). Radiology, however, demonstrated limited predictive value (C-Index = 0.5584), with performance approaching the 0.5 baseline expected from uninformative features. Based on these results, we categorised clinical and histopathology as high-performing modalities for this task, and radiology as low-performing. 

\subsection{Multimodal Model Performance}\

Table 4 presents the results of multimodal fusion using both marginal and joint fusion strategies. Fusing the two high-performing modalities (histopathology and clinical data) yielded a synergistic improvement (C-Index = 0.8348), outperforming either modality in isolation. Conversely, any fusion including the low-performing radiology modality consistently downgraded performance. For example, combining clinical and radiology resulted in a lower C-Index (C-Index = 0.7593) than clinical data alone. Even when radiology was added to the high-performing pair (i.e. clinical, histopathology and radiology), performance dropped (C-Index = 0.8025).

These findings support our hypothesis that multimodal fusion is beneficial when including informative modalities. In contrast, the addition of weaker modalities introduces noise that undermines predictive accuracy. The results highlight the importance of unimodal evaluation prior to fusion, to guide the selective inclusion of modalities.

\begin{table}[t]
\centering
\small
\begin{tabularx}{\columnwidth}{@{}Xcc@{}} 
\toprule
\multirow{2}{*}{\textbf{Modalities}} & \multicolumn{2}{c}{\textbf{Fusion Method}} \\
\cmidrule(lr){2-3}
& \textbf{Marginal} & \textbf{Joint} \\
\midrule
\textbf{Histopathology \& Clinical} & \textbf{0.8210 ± 0.0980}
& \textbf{0.8348 ± 0.0967} \\
Histopathology \& Radiology & 0.7501 ± 0.1166 & 0.7502 ± 0.1389 \\
Clinical \& Radiology & 0.7567 ± 0.1119 & 0.7593 ± 0.1079 \\
Clinical \& Histopathology \& Radiology & 0.8025 ± 0.0985 & 0.7996 ± 0.1062 \\
\bottomrule
\end{tabularx}
\caption{Multimodal fusion results comparing marginal/joint strategies over repeated (10 repeats) 5-fold cross-validation experiments. We present mean C-index $\pm$ standard deviation.}
\label{tab:fusion_results}
\end{table}

\section{Conclusion and Future Directions}
In this study, we investigated the assumption that combining modalities inherently improves performance for computational pathology MDL architectures. We found that integrating high-performing modalities (e.g. clinical data and histopathology) enhances prediction of time-to-BCR, while incorporating low-performing modalities such as radiology degrades performance. This is likely due to the limited ability of preoperative MRI to capture microscopic disease and cellular-level aggressiveness, which histopathology reveals and which ultimately drives recurrence. 

We hypothesised that multimodal gains depend on the predictive quality of individual modalities, and our results validate this. Weak modalities introduce noise rather than complementary information, undermining the predictive power of stronger ones. The success of multimodal fusion is therefore contingent on the discriminative quality of the modalities selected.
While our findings are promising, the small dataset size (n=95) presents a limitation. This restricts statistical power and increases the risk of overfitting, which may affect generalisability to broader patient populations. Future work should focus on curating larger, more diverse datasets and validating this framework across other clinical prediction tasks to confirm its robustness and applicability.

Despite the dataset size limitation, these findings have direct implications for designing multimodal systems in computational pathology. The assumption that `more data is always better' does not hold. Instead, a preceding unimodal analysis is essential to guide the selective inclusion of informative modalities and prevent low-performing ones from introducing noise. This is a principle we believe to be generalisable across diverse clinical applications, and other multimodal domains.

\section{Compliance with Ethical Standards}
This research study was conducted retrospectively using open access human subject data from the Chimera Challenge. Ethical approval was not required.

\section{Acknowledgments}
This work was supported by the Undergraduate Research Support Scheme and the Department of Computer Science at the University of Warwick.

\bibliographystyle{IEEEtran}
\bibliography{refs}
\end{document}